\documentclass[12pt,preprint]{aastex}

\slugcomment{To appear in ApJ, vol. 664}

\shorttitle{Changes in MWC~349A}

\begin{document}


\title{Changes in the Radio Appearance of MWC~349A}


\author{Luis F. Rodr\'\i guez and Yolanda G\'omez}
\affil{Centro de Radioastronom\'\i a y Astrof\'\i sica, Universidad
Nacional Aut\'onoma de M\'exico, Morelia 58089, M\'exico}
\email{l.rodriguez, y.gomez@astrosmo.unam.mx}

\and

\author{Daniel Tafoya\altaffilmark{1}}
\affil{Harvard-Smithsonian Center for Astrophysics,
60 Garden Street, Cambridge, MA 02138, USA}
\email{dtafoya@cfa.harvard.edu}


\altaffiltext{1}{Centro de Radioastronom\'\i a y Astrof\'\i sica, Universidad
Nacional Aut\'onoma de M\'exico, Morelia 58089, M\'exico}


\begin{abstract}
We present new sensitive, high angular resolution 1.3, 2, and 6 cm observations of
the continuum emission from the peculiar emission-line star MWC 349A, made with the
Very Large Array. This radio emission is believed to
originate in an ionized flow produced by the photoevaporation of a disk that surrounds
the star. We determine for the first time the proper motion of this source,
which is consistent with that expected for the location of the
source in the galaxy.
Our analysis of the images, that include the new observations
as well as archive data covering a time interval of more than 20 years,
indicates that the appearance of MWC~349A has been systematically changing over time.
The well-defined ``hourglass'' shape that characterized the 2 and 1.3 cm appearance of
the source in the early 1980's has disappeared to be replaced by a more
``square'' shape.
We discuss if these changes can be accounted for by precession of the
MWC~349A disk or by intrinsic changes in the parameters of the disk,
but could not reach a satisfactory explanation.

\end{abstract}





\section{Introduction}

MWC~349A is the brightest thermal radio continuum star in the 
centimeter domain (Braes, Habing, \& Schoenmaker 1972). At a 
distance of 1200 pc it has a bolometric luminosity
of 30,000 L$_\odot$ (Cohen et al. 1985) and exhibits an unusually 
slow and dense stellar wind with a velocity of $\sim$50 km s$^{-1}$ 
(Hartmann, Jaffe, \& Huchra 1980; Altenhoff et al. 1981; Tanaka et al. 1985). 
It is believed that the massive, slow outflow originates
in the surfaces of a neutral photoevaporating disk (Hollenbach et al. 1994;
Lugo, Lizano, \& Garay 2004; Tafoya, G\'omez, \& Rodr\'\i guez 2004). 
This disk has been inferred by: i) the presence of a double-peaked profile 
at optical, IR and radio recombination
maser lines (Hamann \& Simon 1986; Planesas et al. 1992); 
ii) the rotation of the envelope around MWC 349A (Rodr\'\i guez \& Bastian 1994); 
iii) the presence of a dark lane in the equatorial plane of the source 
(White \& Becker 1985; Tafoya, G\'omez \& Rodr\'\i guez 2004), and 
iv) observations in the near infrared that show a structure aligned 
in the direction of the equatorial dark lane (Danchi et al. 2001; Hofmann et al. 2002).
The high angular resolution  7 mm observations (36 mas) made by Tafoya et al. (2004),
show clearly the presence of a dark lane due to the lack of free-free emission from 
the neutral disk. These authors estimate a lower limit for the electron density
in the wind
of $n_{e}\geq1.4\times 10^{7}$ cm$^{-3}$ at a distance of 50 AU from the
star.
This lower limit is consistent with the predicted electron density necessary to 
explain the H30$\alpha$ maser recombination line (Mart\'\i n-Pintado et al. 1989; 
Gordon 1992). MWC 349A is the only known source to exhibit maser amplification
in its mm and sub-mm recombination lines.

The radio continuum emission from MWC~349A shows an extended bipolar
structure with a waist that gets narrower as we observe at high frequencies (Tafoya et al. 2004). 
The radio spectrum is consistent with that expected for a
biconical thermal wind that expands at constant velocity (e. g. Olnon 1975; Tafoya et al. 2004).
The radio continuum emission also shows a faint structure to the west that 
suggests an interaction between the winds of MWC 349A and 
that of its companion MWC 349B (Cohen et al. 1985; Tafoya et al. 2004).   
MWC 349B has been classified as a B0 III star (Cohen et al. 1985).

Variability in the optical regime and in the millimeter maser recombination
lines has been found toward
MWC 349A (Bergner et al. 1995; Thum et al. 1992). In particular, Jorgenson 
et al. (2000) reported a red light curve covering the years 1967-1981,
that suggests a period of 9 yr and a probable amplitude variation of $\pm$0.4 mag. 
The study of Tafoya et al. (2004) suggested the presence of
variation at 6 cm over the 14.5 years separating the two images
analyzed. In this paper we report new 1.3, 2, and 6 cm observations
as well as an extensive analysis of archive data to investigate
possible time variations in more detail.

\section{Observations}


The new 1.3, 2, and 6 cm observations were made in the A configuration
of the VLA of the NRAO\footnote{The National Radio 
Astronomy Observatory is operated by Associated Universities 
Inc. under cooperative agreement with the National Science Foundation.},
during 2004 November 15. 
The data were edited and calibrated following
the standard VLA procedures. 
The absolute amplitude
calibrator was 1331+305 (with adopted flux densities of 2.52, 3.46, and 7.49 Jy
at 1.3, 2, and 6 cm respectively)
and the phase calibrator was 2007+404 (with bootstrapped flux densities
of 3.20$\pm$0.11, 3.18$\pm$0.02, and 2.53$\pm$0.03 Jy
at 1.3, 2, and 6 cm respectively). 
The effective bandwidth of the observations
was 100 MHz. All data were self-calibrated in phase. 

For the discussion of time behavior of the source, we also used extensively data from the
VLA archives, that are discussed in the following sections.




\section{Absolute Astrometry}

We used the four data sets obtained in the highest angular
resolution A configuration at 2 cm that
are listed in Table 1 to measure the
proper motions of MWC 349A. These four data sets are appropriate for this
determination since they span a large time baseline (1983.83 to 2004.87),
have high angular resolution ($\sim 0\rlap.{''}1$), and were all taken
with the same nearby phase calibrator, 2007+404.  
The position used for 2007+404 in the four epochs was updated to
the most recent position given by the VLA calibrator manual,
$\alpha(2000) = 20^h~07^m~44\rlap.^s944851;~\delta(2000) = 
+40^\circ~29{'}~48\rlap.{''}604140$.
We fitted the position of MWC 349A
with two techniques: 1) doing a least-squares fit
of the source to a Gaussian ellipsoid 
using the task JMFIT of AIPS and 2) determining the
offsets with respect to the 2004.87 epoch by minimizing the
$\chi^2$ expression given by

$$\chi^2 = \sum_{x,y} [I(x-x_o,y-y_o) - I_{ref}(x,y)]^2,$$

\noindent where $I$ is the intensity of the image being aligned, $I_{ref}$ is 
the intensity of the reference
image (in this case, the most recent image, taken in 2004.87), and
$x_o$ and $y_o$ are the offsets. The sum is made over a rectangular box
containing all the detectable emission from the source. The centroid
position of the 2004.87 epoch image was determined to be
$\alpha(2000) = 20^h~32^m~45\rlap.^s5284;~\delta(2000) = 
+40^\circ~39{'}~36\rlap.^{''}623$ from the task JMFIT.

Both methods gave consistent results and we finally used the $\chi^2$ method
that gives the proper motions shown in Fig. 1. The positional
errors at a given epoch were found to be of
order 10 mas. A least-squares fit to
these data gives average proper motions of:

$$\mu_\alpha cos(\delta) = -3.1 \pm 0.5~mas~yr^{-1},$$

$$\mu_\delta = -5.3 \pm 0.5~mas~yr^{-1}.$$

To our knowledge, MWC~349A has no previously reported proper motions.
Clemens \& Argyle (1984) tried to measure its proper motion from optical images,
but their errors ($\sim 8~ mas~yr^{-1}$) were about
an order of magnitude larger than ours and 
this precluded a determination.
The proper motions found by us for MWC~349A
are similar to those measured by Hipparcos
(Perryman et al. 1997) for 
the star P Cygni, $\mu_\alpha cos(\delta) = -3.5 \pm 0.4~mas~yr^{-1};
\mu_\delta = -6.9 \pm 0.4~mas~yr^{-1}$. P Cygni is located at $3\rlap.^\circ9$
in the plane of the sky from MWC~349A. We tried to use the determined
proper motions of MWC~349A to set limits to its distance. Unfortunately,
the proper motions in these galactic coordinates for objects at a few kpc from the
Sun are relatively insensitive to the distance and
we could not improve on what is known.

\section{2 cm Flux Density as a Function of Time}

As discussed before, the results of Tafoya et al. (2004) suggested
small variations, of order 1\%, between the two 6 cm images
taken with a time separation of 14.5 years and analyzed by them.

We first analyzed if there is evidence in the archival data
accumulated over the last two decades for significant
variations in the total continuum flux density of MWC~349A. For this analysis
we used four data sets taken at 2 cm in the lowest angular resolution D configuration
of the VLA. These observations of low angular resolution are more appropriate for
measuring the total flux density of the source since they are less affected by
the lack of short spacings or by phase noise produced by poor weather than 
those taken with higher angular resolution. In Table 2, we list the
parameters of the observations and the flux densities determined. All values
are consistent with a 2 cm flux density of 384$\pm$6 mJy, suggesting that the source
has not suffered significant changes in its total 2 cm flux density over the last 20 years. 
This flux density is consistent at the 15\% level with the power law
fits given by Altenhoff et al. (1981) and
Tafoya et al. (2004) to the continuum spectrum of MWC~349A.
This lack of variability in the radio is previously
known for MWC 349A,
which allows its use as a flux calibrator
in interferometers such as BIMA (e. g. Chen et al. 2006) and IRAM 
(e. g. Pi\'etu et al. 2006).
We will then assume that there is no variation in the 
total flux density of the source
and for the morphology
comparisons we will adjust the flux density scale of the observations made at different
epochs to the flux density values given by the power law fit of Tafoya et al. (2004).

\section{Analysis of Morphology as a Function of Time}

\subsection{2 cm Images}

For this analysis, we used the same four high angular resolution data sets
used for the absolute astrometry (see Table 1). To allow an easier comparison, we
made images with the same circular restoring beam of $0\rlap.{''}11$ of
half power full width. We consider this procedure adequate because the
dimensions of the
true synthesized beams of the observations differ only slightly
from this value (see Table 1). These images are shown in Fig. 2.
When comparing the images as a function of time, a remarkable variation
is observed. While in 1983, the radio source had a well defined ``waist''
(that gave the source its ``hourglass'' shape)
running approximately in the east-west direction,
over the years this structure progressively disappeared, until by 2004 the source showed a
more or less ``square'' morphology. These changes are also evident in the
difference images, made from subtracting to the 2004 image the other epochs.
In these difference images we see that the emission in the original north and south lobes
diminished with time while that in the ``waist'' of the nebula increased.
We consider these results reliable since all four 2 cm databases used have similar
(u,v) coverage, ranging from 40 to 1,500 k$\lambda$. 

\subsection{1.3 cm Images}

For this analysis, we used the three data sets presented in Table 3.
The 1.3 cm data is in general noisier than that at 2 cm and the variation is not
as clear as at the latter wavelength. 
However, as can be seen in Fig. 3, the same trend is present in the 1.3 cm
images, with the ``waist'' of the nebula disappearing with time. 

\subsection{6 cm Images}

For this analysis, we used the three data sets presented in Table 4.
Perplexingly, the time variation is different than that seen at 2 and
1.3 cm and consistent with the variation reported at 6 cm by Tafoya et al.
(2004) using only the first two epochs (see Fig. 4). The new 2004 observations confirm
the trend observed by Tafoya et al.. that is,
the difference images clearly show a negative broken "ring"
(with openings to the north and south) at a radius of $\sim0\rlap.{''}5$. 
This negative broken ring becomes more pronounced in the 2004.87-1982.42
image than in the 2004.87-1996.93, as expected
for an effect that increases with time.
The fact that the 6 cm images show a different behavior to that found in the 2 and
1.3 cm images is not in conflict with what is known of this source:
in an ionized wind such as MWC 349A there is a radio
``photosphere'' that goes approximately as $\nu^{-0.7}$, implying that at 6 cm we are observing
a region about 2 to 3 times larger that at 2 and 1.3 cm. Then, the behaviors observed at
different wavelengths do not have to coincide. In what remains of the paper
we will concentrate our efforts in trying to account for the stronger morphology variations
detected at 2 and 1.3 cm.

\section{Interpretation}

\subsection{Precession of the MWC 349A Disk?}

One possibility is that the disk associated with MWC 349A has experienced significant
precession over the 20 years of the study. In this possible explanation we will assume
that the disk has suffered no intrinsic changes and that the observed 
differences are
just a result of a change in orientation with respect to the observer.

To test if a precession of the disk could account for the changes, we made a
model following the assumptions of White \& Becker (1985) and adding to the
model the possibility of tilting the plane of the disk in the direction
of the observer at a given inclination angle. This model is known to reproduce
only approximately the appearance of the source. In particular, as
emphasized by
White \& Becker (1985), it does not reproduce the emission at large radii in the plane of the disk.
Our comparison has then to be considered only qualitative.
Nevertheless, analysis of Fig. 5 suggests that precession can indeed
account qualitatively for the observed changes, in the sense that the nebula becomes progressively
more square with increasing inclination and that
from the point of
view of the observer it is found that flux is ``shifted'' from the lobes to the plane of the disk. 

However, is precession with such a small timescale (decades) possible?
One could attribute the disk precession to the
presence of the companion star MWC 349B. We will assume masses
of $M_p = 30~M_\odot$ for MWC 349A (the primary star) and
of $M_s = 18~M_\odot$ for MWC 349B (the secondary star),
as estimated from the results of Rodr\'\i guez \& Bastian (1994)
and Cohen et al. (1985). Assuming a distance of 1200 pc, we
estimate that the stars are separated in the plane of the sky
by a distance of
$D = 2900~AU$ and that the outer radius of the ionized disk at 2 cm is
$R = 360~AU$. Using the formulation of Terquem et al. (1999), the
period of the precession of the disk (present around the primary star)
is given by

$$P_p \simeq 1.3\times10^1~\Biggl({{M_p} \over {M_s}} \Biggr)~
\Biggl({{D} \over {R}} \Biggr)^3~\Biggl({{R^3} \over {G~M_p}} \Biggr)^{1/2}, $$

\noindent where $G$ is the gravitational constant.  
From the values given above, we estimate $P_p \simeq 2.3 \times 10^6$ years,
far much larger that the timescale covered by the observations.
We conclude that, although precession could explain the
observed variations, the timescales expected from the effect of
MWC 349B are just too large.

It is possible that if another star much closer to MWC349A is present, the precession could
be faster. In the case of the source NGC~7538 IRS1, Kraus et al. (2006)
present evidence of disk precession with a period of $\sim$280 years, comparable
with the timescales required for a precession model in MWC 349A.
Kraus et al. (2006) propose as a possible triggering mechanism for 
precession in NGC~7538 IRS1 the non-coplanar tidal interaction of an (undiscovered) 
close companion with the circumbinary protostellar disk. 
Hofmann et al. (2002) have suggested that
MWC 349A could be a close binary star, but there is no direct evidence for this.
Furthermore, the estimated period for rotation of the outer
radius of the MWC 349A disk
is of order $10^3$ years and it does not seem plausible to have precession
scales on much shorter scales.

\subsection{Intrinsic Changes in the Disk?}

Another possibility to explain the observed changes is to have the
disk itself experience significant changes. We do not know the structure and
physical conditions of the disk with sufficient precision to model these possible
changes and will only discuss if the expected timescales for variation
are consistent with a few decades. 

The kinematical timescale of the MWC 349A disk can be estimated by dividing the radius of
the disk (360 AU) over the speed of the wind ($\sim$50 km s$^{-1}$), to obtain a
timescale of about 35 years. A variation in this timescale could be related, for example,
to the wind rapidly clearing the disk or to the photoevaporation occuring
very rapidly. 

Another timescale of relevance is that given by the electron recombination time, which for
an electron density of $10^5$ cm$^{-3}$ (as expected in the outer parts
of the disk) and an electron temperature of
$10^4$ K, gives about one year. A variation in this timescale could be related, for example,
to the recombination or ionization of significant
volumes of gas associated with the disk, presumably produced by changes
in the central star. 

Keto (2007, private communication) has used his model of a
rotating accretion flow into a forming
massive star (Keto 2007) to find that increases of order
10-30\% in the electron density of the flow can produce morphological
changes that resemble the changes observed in MWC 349A over  
the last two decades.

We conclude that, based on the timescales involved,
the possibility on intrinsic changes in the disk seems more
plausible than that of precession. To obtain a qualitative impression of the
appearance changes expected from changes in the disk, we have again used the model of
White \& Becker (1985) and, for an inclination angle of $10^\circ$ with respect to
the observer, we have varied the critical angle $\theta_c$ above which the gas
is fully ionized. The critical angle is equivalent to the flaring angle
of the disk. These variations could be produced by very rapid photoevaporation
of the surface of the disk. As can be seen in Fig. 6, again we have
that flux from the lobes ``moves''
to the waist of the nebula and the appearance becomes squared, as we diminish
$\theta_c$. 

However, as in the case of precession, the expected timescales for significant
photoevaporation of the disk are much larger than a few decades. The mass of the
MWC 349A disk is not known, but Danchi et al. (2001) give upper limits of
a few solar masses. Tafoya et al. (2004) give a mass loss rate in ionized gas of
$5.0 \times 10^{-6}~M_\odot~yr^{-1}$. Assuming a mass of 1 $M_\odot$ for the disk, we
find a photoevaporation timescale of $2 \times 10^5~yr$.
An even larger timescale, of several million years, is given by
Yorke \& Weltz (1996) from models of the evolution of photoevaporating disks around
early B type stars.

\section{Conclusions}

We have analyzed new as well as archive continuum VLA data of the peculiar emission-line star MWC 349A
covering more than 20 years of time.
We determined for the first time the proper motions of this source,
which are consistent with those expected for its location 
in the galaxy.

The appearance of MWC~349 has been systematically changing over time
and we discuss if those variations are due to precession of the disk or
to intrinsic changes in the source. Although these types of variations
produce changes qualitatively similar to those found in the
observations, the timescales expected for them are much larger than
the two decades comprised in the observations. We could not reach a
satisfactory explanation for the
variations and additional detailed theoretical modeling and
continuous monitoring are needed to advance in
our understanding of this source.

\acknowledgments

LFR and YG acknowledge the support
of CONACyT, M\'exico and DGAPA, UNAM.



{\it Facilities:} \facility{VLA}

\clearpage



\begin{figure}
\epsscale{1.10}
\plottwo{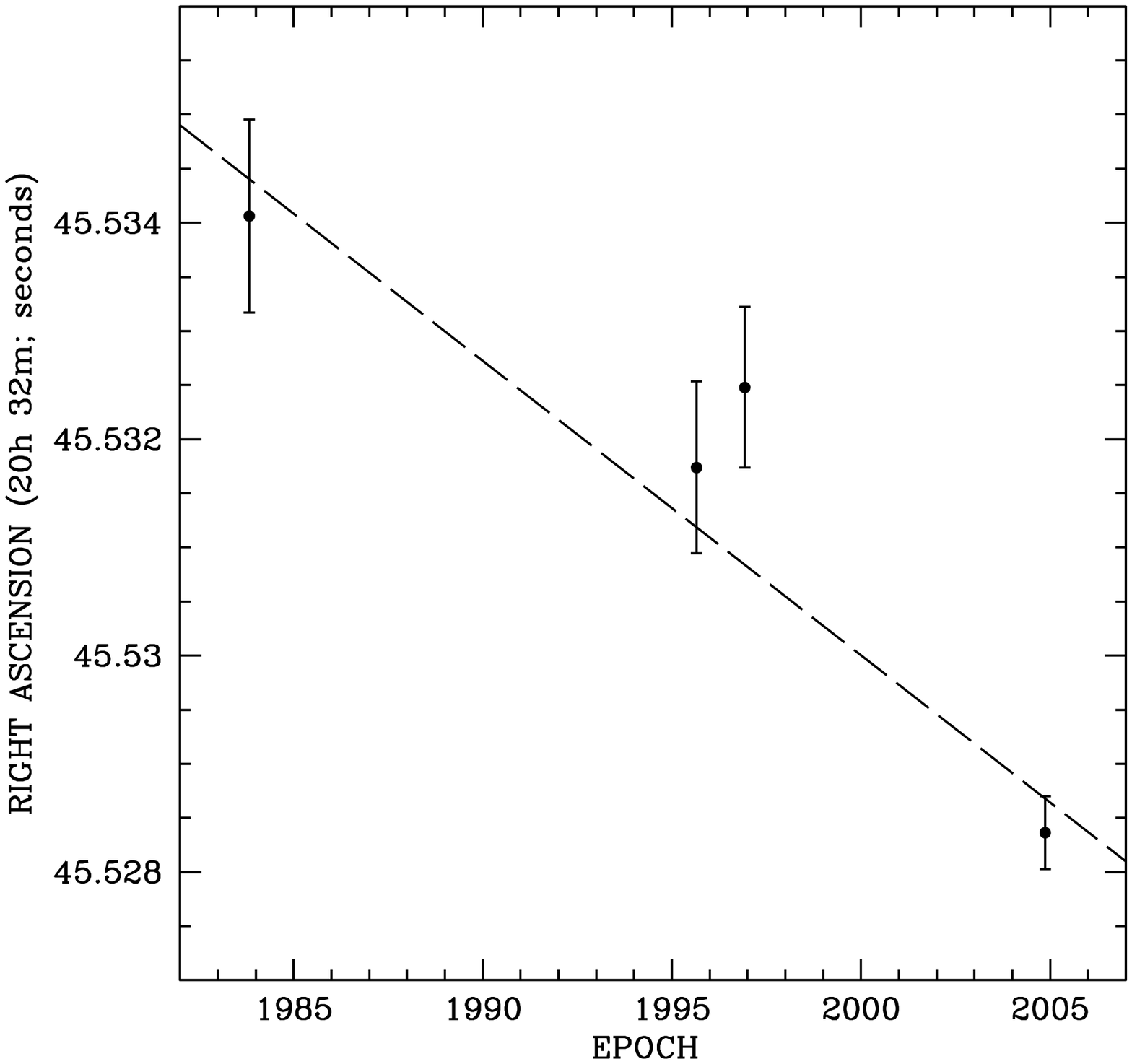}{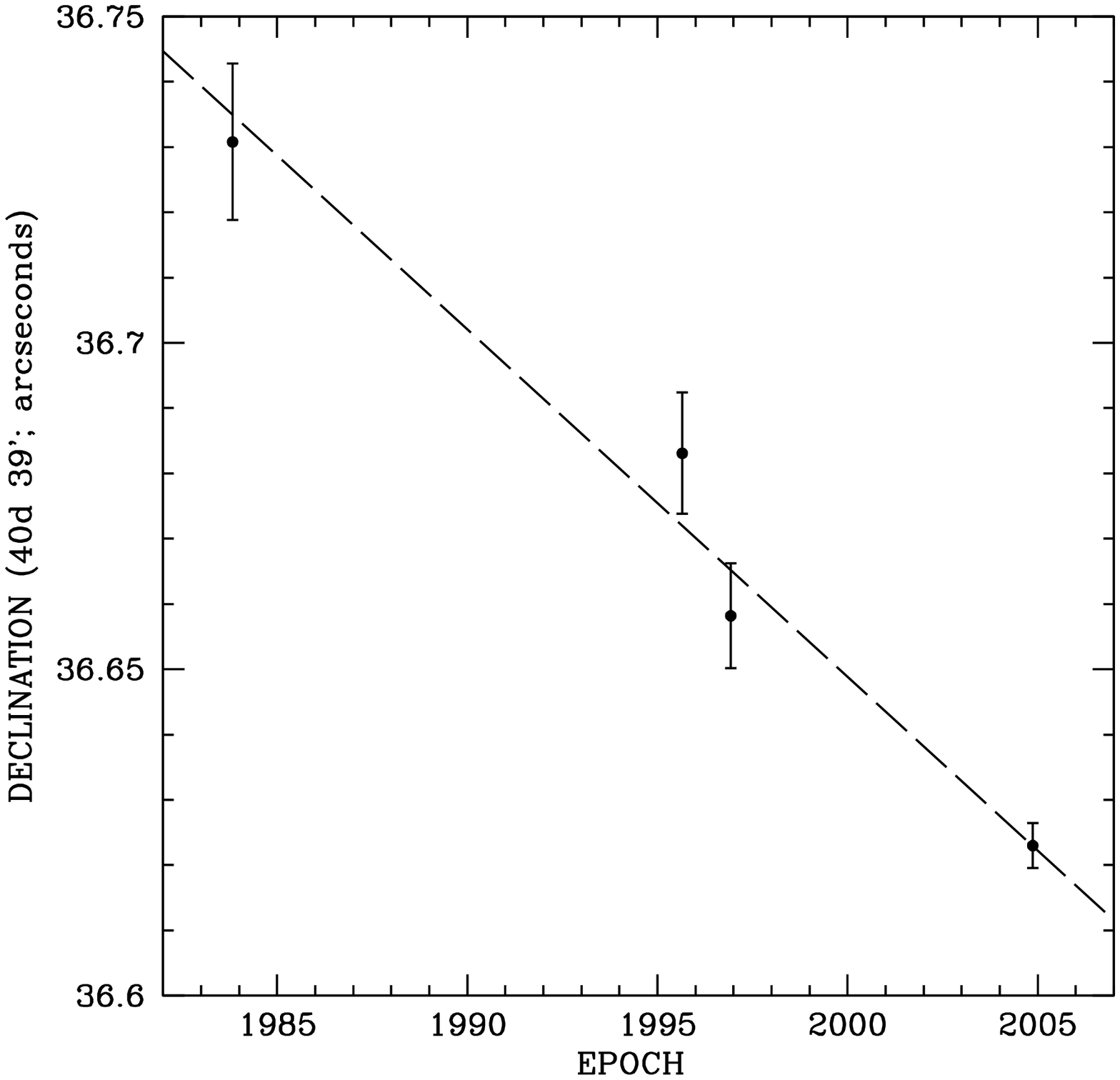}
\caption{Positions in right ascension (left)
and declination (right) for MWC~349A as a function
of epoch. These positions
are obtained from the four epochs listed in Table 1. The
dashed lines are the least-squares fits to the data, that
give proper motions $\mu_\alpha cos(\delta) = -3.1 \pm 0.5~mas~yr^{-1};~
\mu_\delta = -5.3 \pm 0.5~mas~yr^{-1}.$
\label{fig1}}
\end{figure}

\clearpage

\begin{figure}
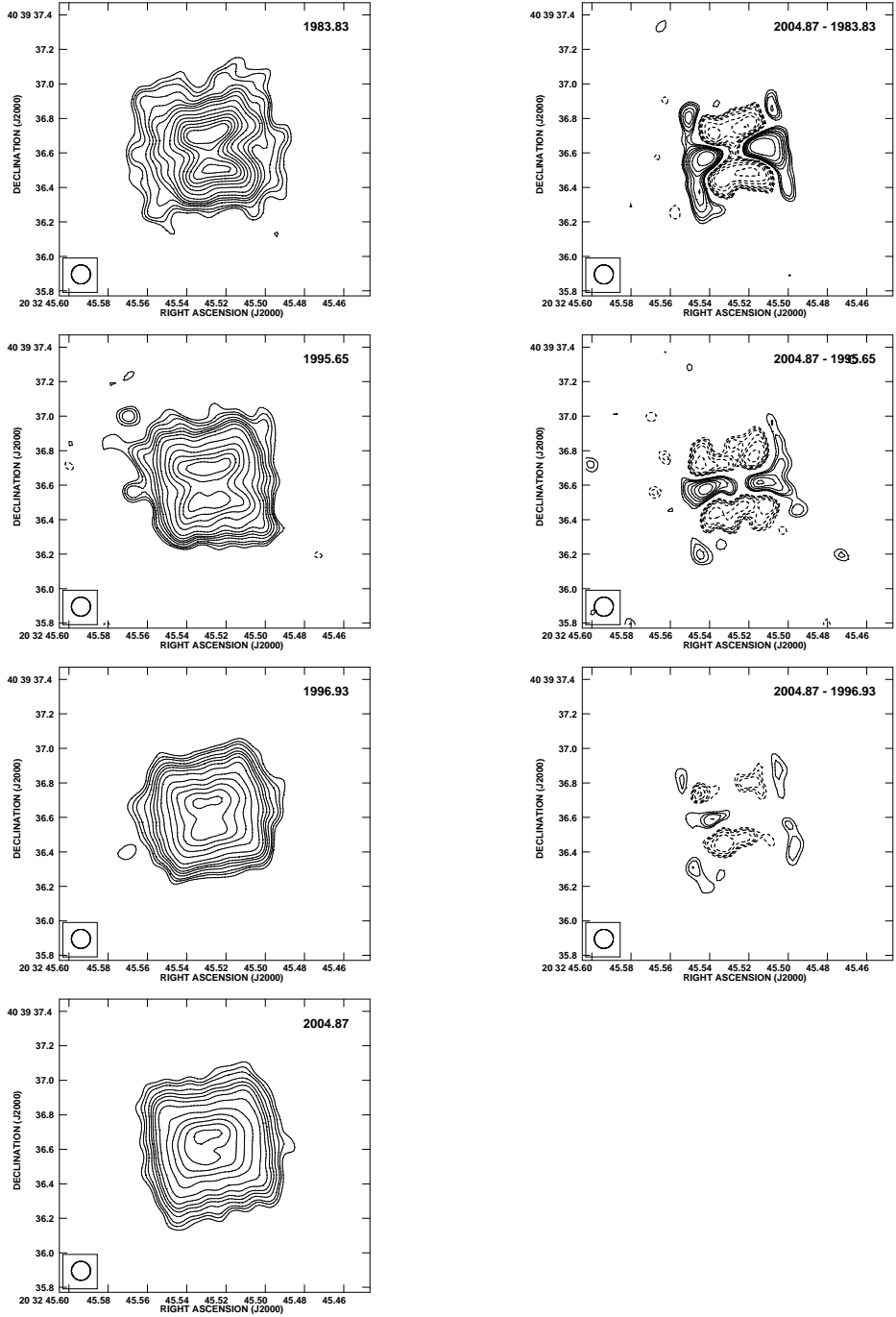

\epsscale{0.70}
\begin{center}

\plottwo{f2a.eps}{f2e.eps}

\plottwo{f2b.eps}{f2f.eps}

\plottwo{f2c.eps}{f2g.eps}

\plottwo{f2d.eps}{f2h.eps}

\end{center}

\caption{(Left) Contour images of the 2 cm emission from MWC~349A for four
different epochs. (Right) Contour plots of the 2 cm difference images.
Contours are
-20, -15, -12, -10, -8, -6, -5,-,4, 4,
5, 6, 8, 10, 12, 15, 20, 30, 40, 50, 60, 70, and 80 
times 0.30 mJy beam$^{-1}$.
The half power contour of the restoring beam ($0\rlap.{''}11 \times 0\rlap.{''}11$)
is shown in the bottom left corner. The images of the first
three epochs are corrected for proper motion
and aligned to the 2004.87 position of MWC~349A.
\label{fig2}}
\end{figure}

\clearpage

\begin{figure}
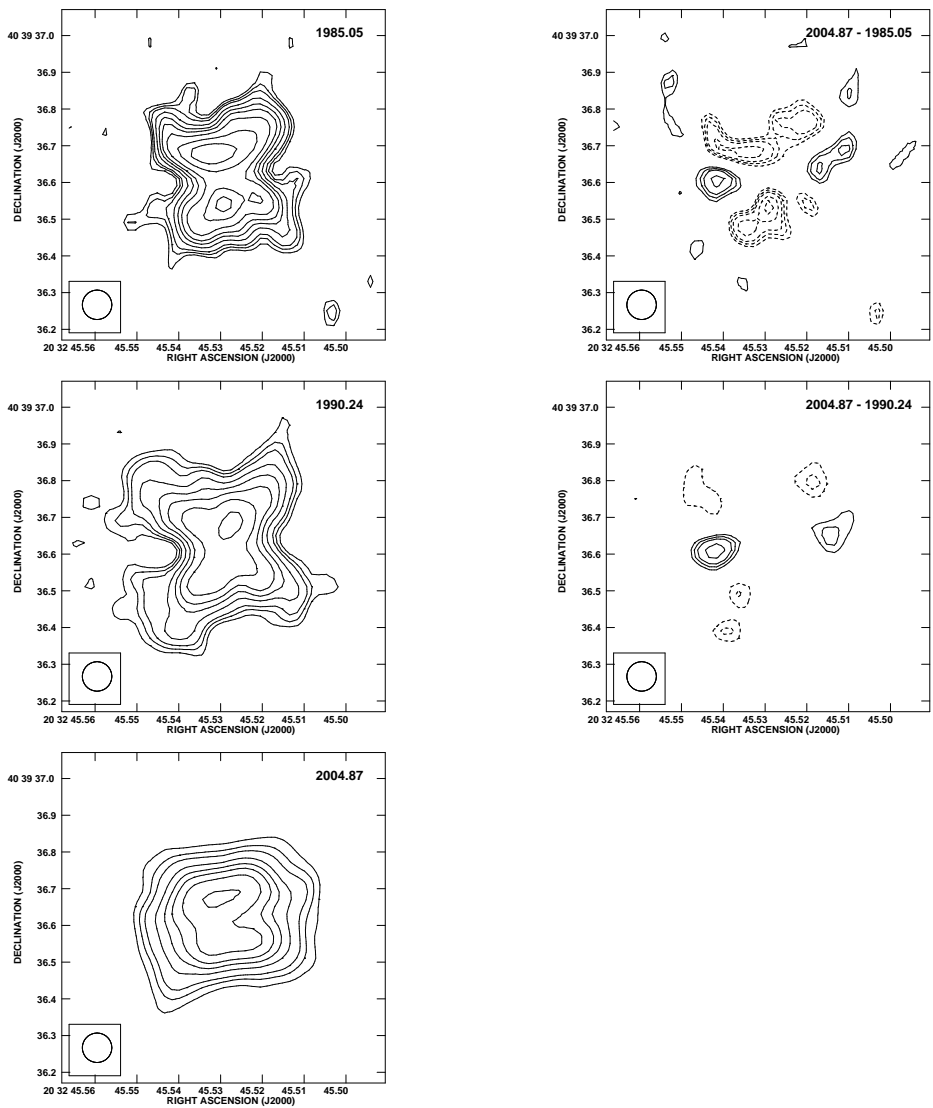

\epsscale{0.70}
\begin{center}

\plottwo{f3a.eps}{f3d.eps}

\plottwo{f3b.eps}{f3e.eps}

\plottwo{f3c.eps}{f3f.eps}

\end{center}

\caption{(Left) Contour images of the 1.3 cm emission from MWC~349A for three 
different epochs. (Right) Contour plots of the 1.3 cm difference images.
Contours are
-10, -8, -6, -5,-,4, 4,
5, 6, 8, 10, 12, 15, 20, and 25
times 1.2 mJy beam$^{-1}$.
The half power contour of the restoring beam ($0\rlap.{''}08 \times 0\rlap.{''}08$)
is shown in the bottom left corner. The images of the first
three epochs are corrected for proper motion
and aligned to the 2004.87 position of MWC~349A.
\label{fig3}}
\end{figure}

\clearpage

\begin{figure}
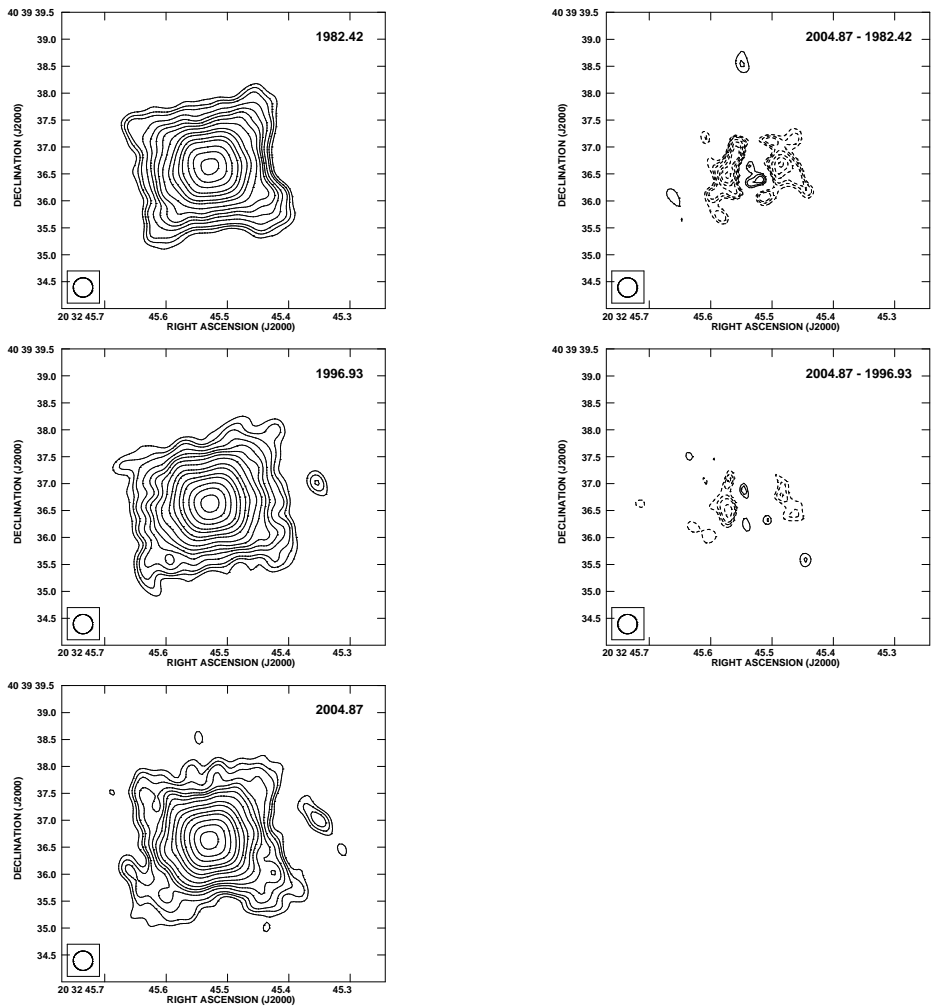

\epsscale{0.70}
\begin{center}

\plottwo{f4a.eps}{f4d.eps}

\plottwo{f4b.eps}{f4e.eps}

\plottwo{f4c.eps}{f4f.eps}

\end{center}

\caption{(Left) Contour images of the 6 cm emission from MWC~349A for three
different epochs. The feature to the west of MWC 349A is the
interaction zone of the stellar winds located between MWC~349A and MWC~349B
that is discussed in detail by Tafoya et al. (2004).
(Right) Contour plots of the 6 cm difference images.
Contours are
-20, -15, -12, -10, -8, -6, -5,-,4, 4,
5, 6, 8, 10, 12, 15, 20, 30, 40, 60, 80, 120, 160, and 200
times 0.10 mJy beam$^{-1}$.
The half power contour of the restoring beam ($0\rlap.{''}36 \times 0\rlap.{''}36$)
is shown in the bottom left corner. The images of the first
three epochs are corrected for proper motion
and aligned to the 2004.87 position of MWC~349A.
\label{fig4}}
\end{figure}

\clearpage

\clearpage

\begin{figure}
\epsscale{1.00}
\begin{center}

\plottwo{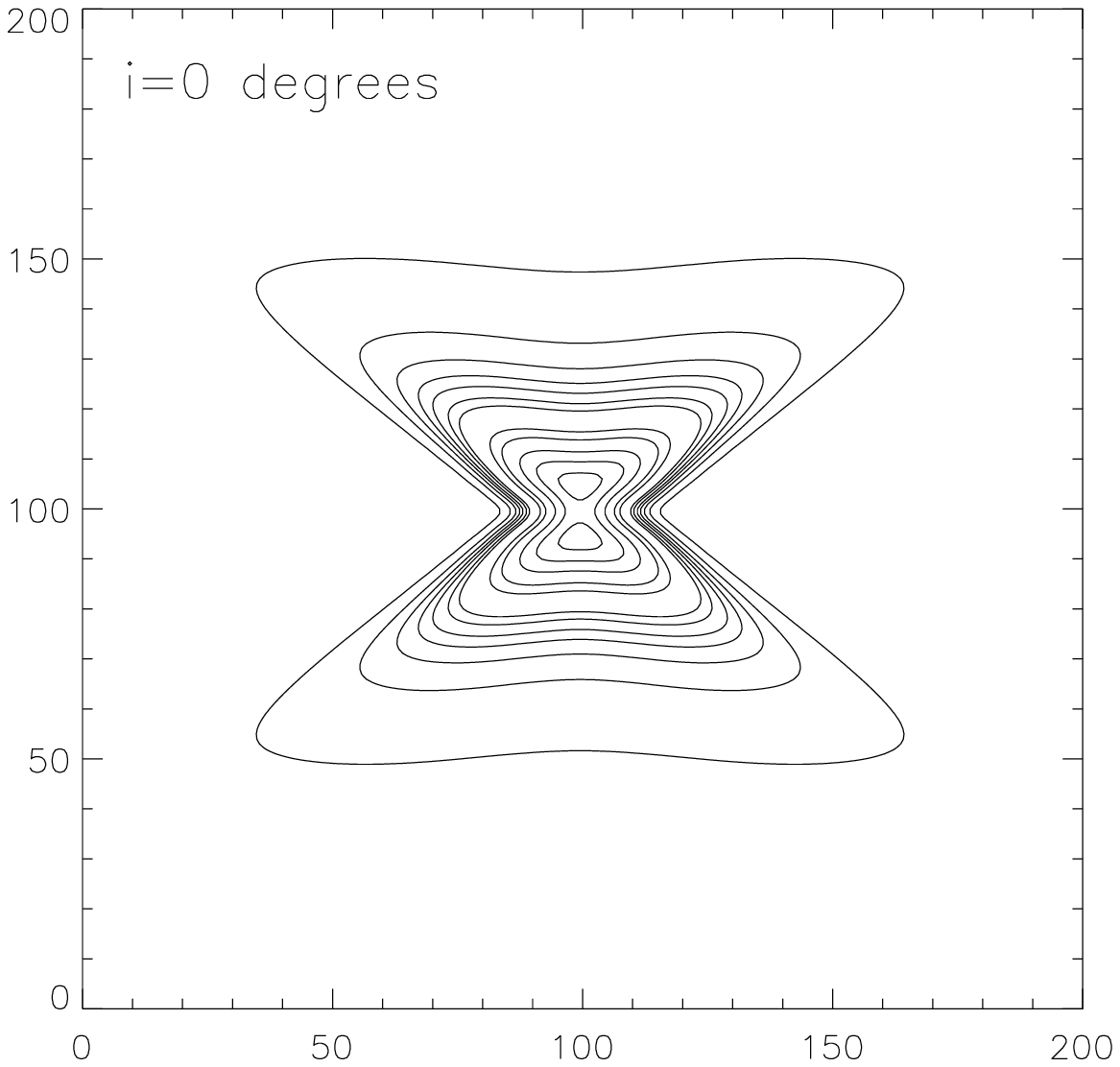}{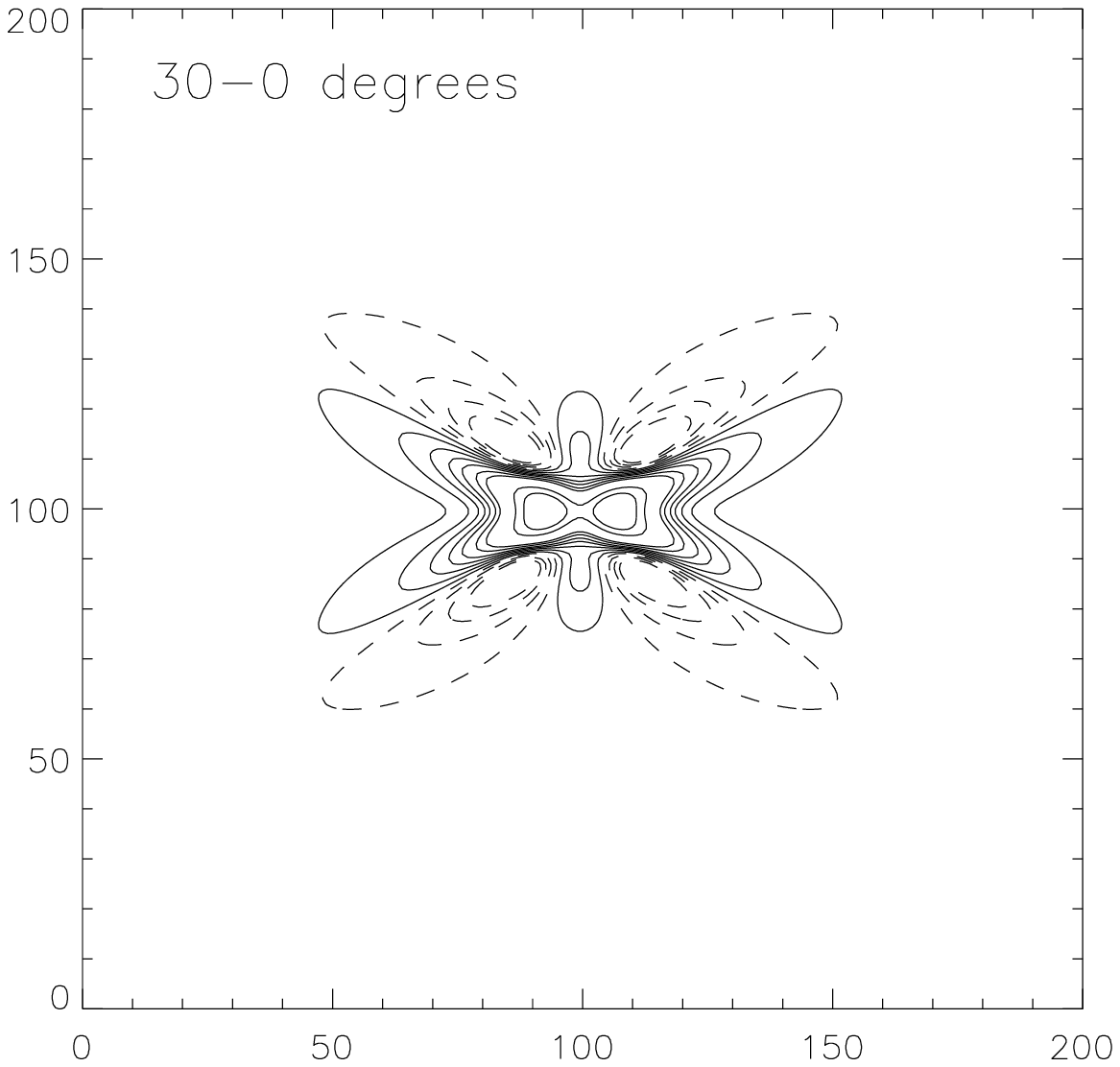}

\plottwo{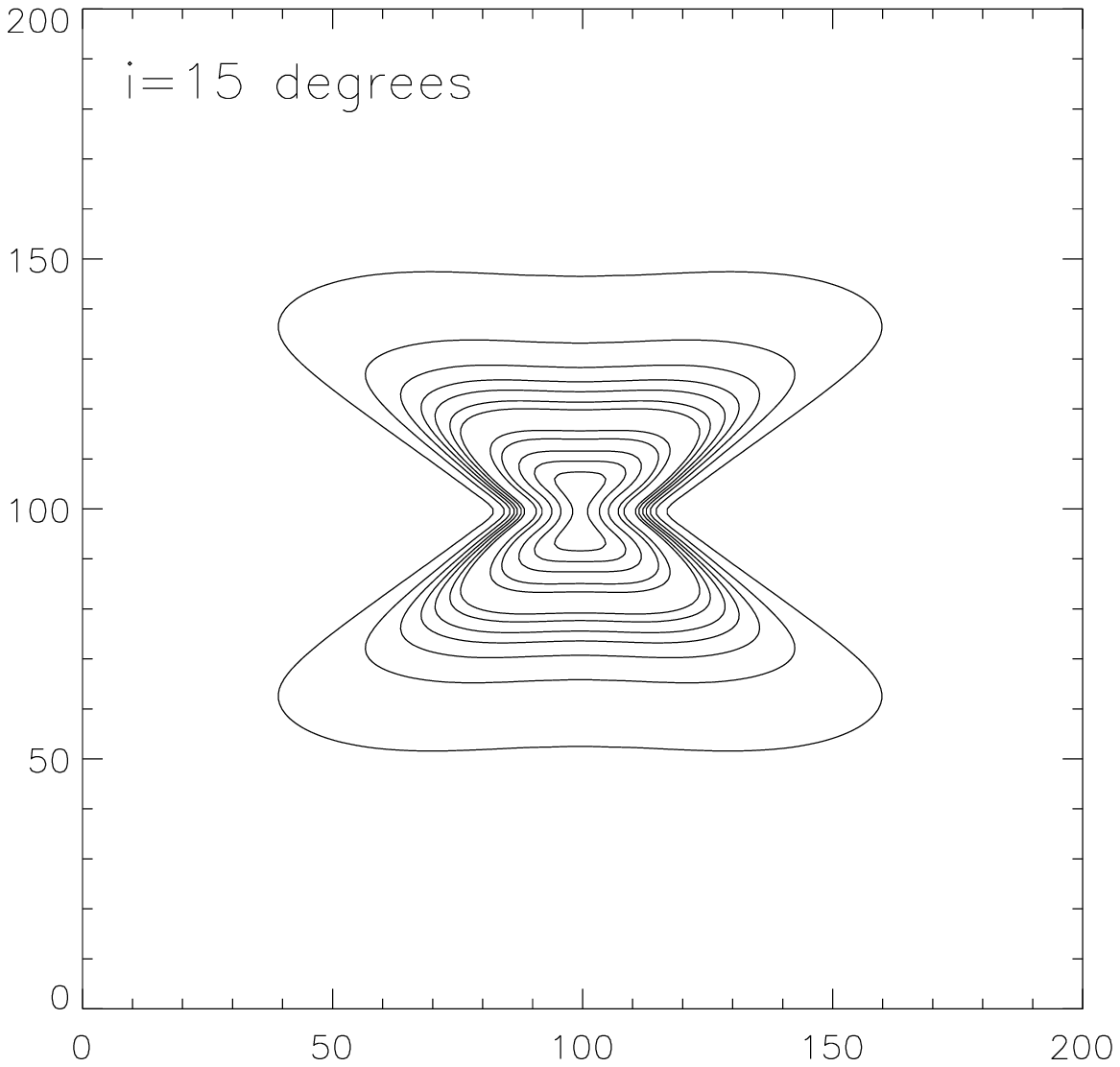}{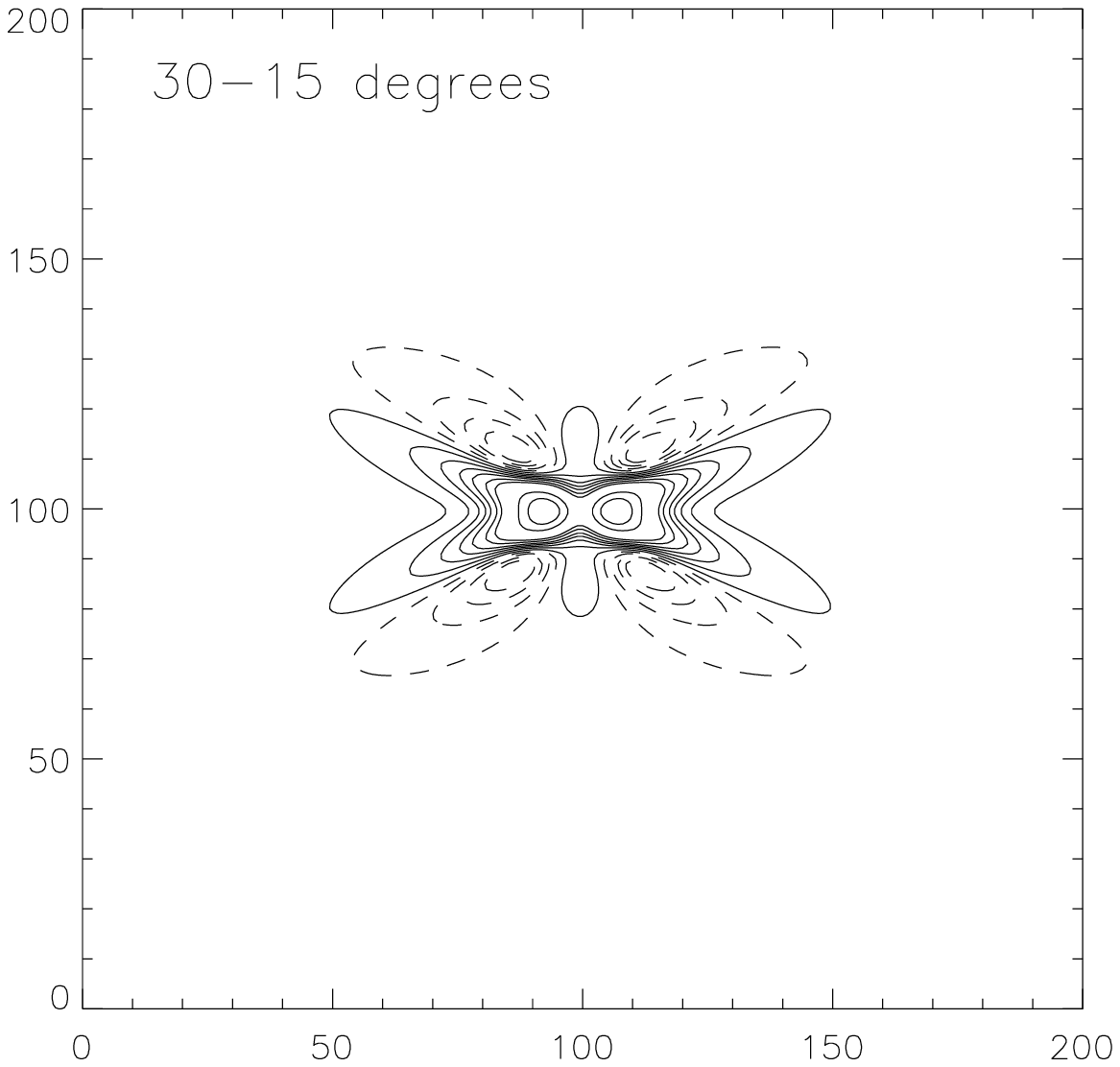}

\plottwo{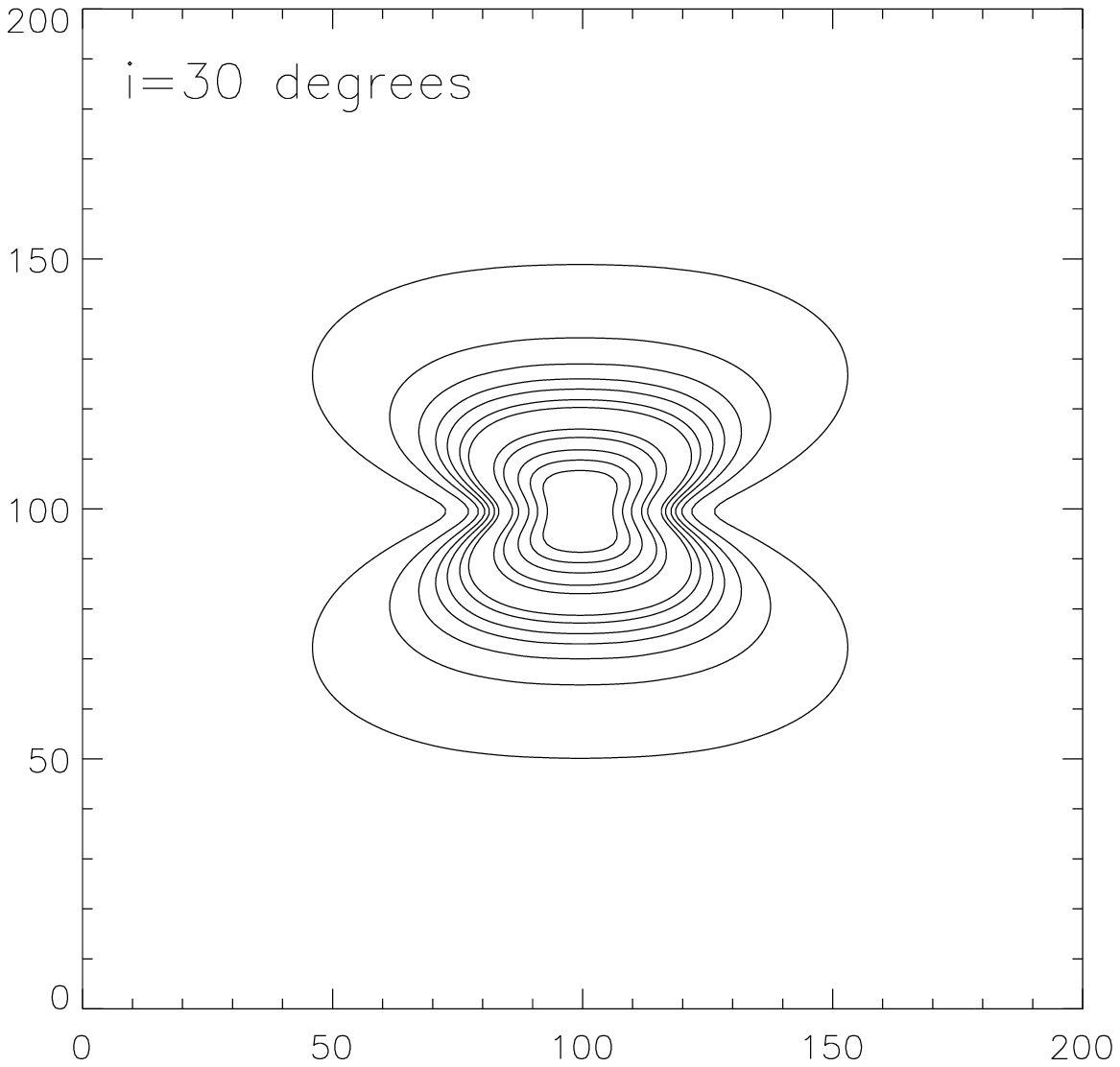}{f5f.eps}

\end{center}

\caption{(Left) Contour images of the precession model described in the text.
The inclination angle of the plane of the disk of the model with respect to
the line of sight is indicated in each figure.
(Right) Contour plots of the model difference images.
Contours are
-100,-80, -60, -40, -30, -15, -12, -9, -7, -5, -3, -1,
1, 3, 5, 7, 9, 12, 15, 30, 40, 60, 80, and  100
percent the peak value of the model with
inclination angle of $0^\circ$.
The $x$ and $y$ axes of the models are in units of $0\rlap.{''}01$ and the images are smoothed to an
angular resolution of $0\rlap.{''}1$.
\label{fig5}}
\end{figure}

\clearpage

\begin{figure}
\epsscale{1.00}
\begin{center}

\plottwo{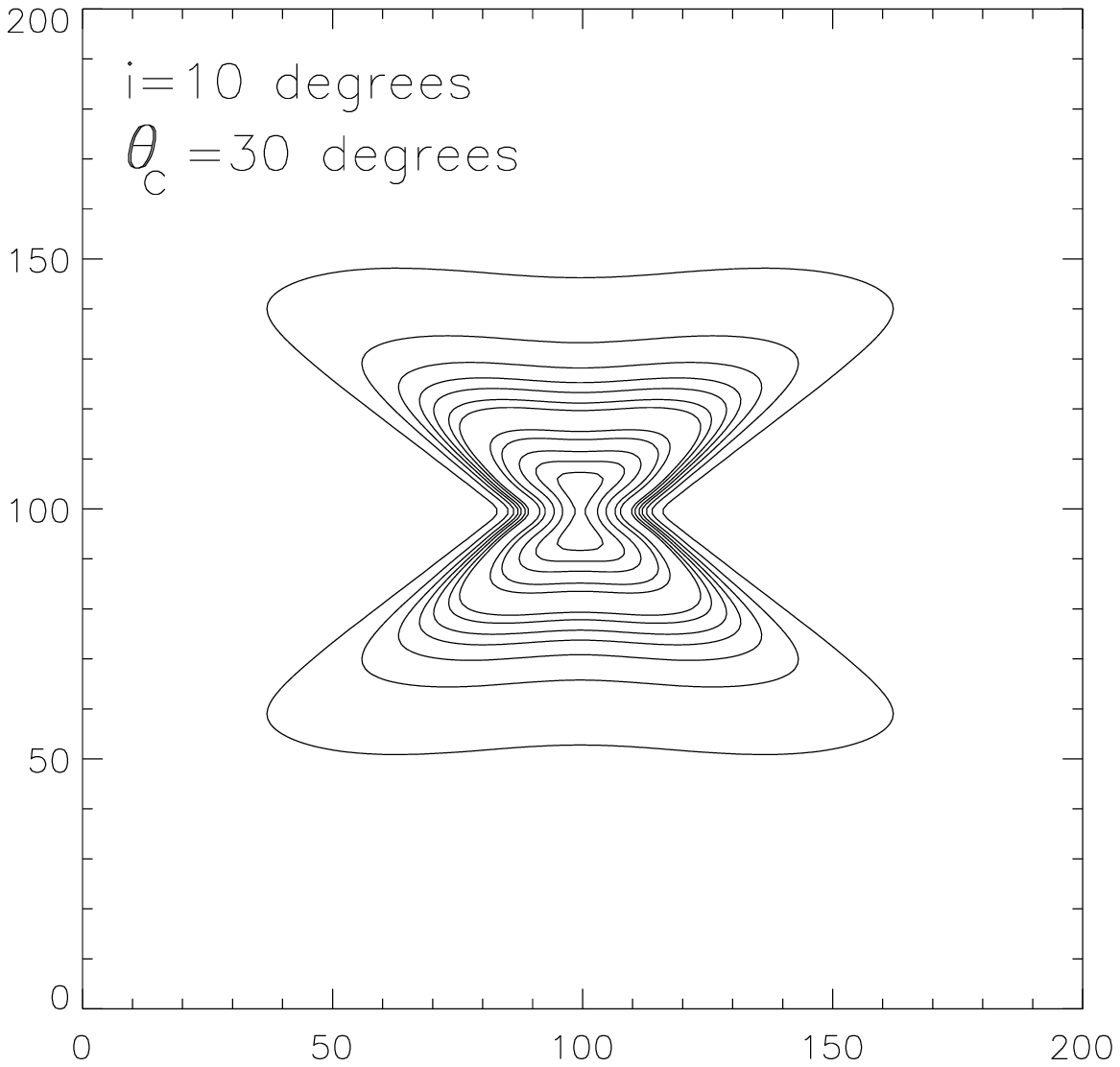}{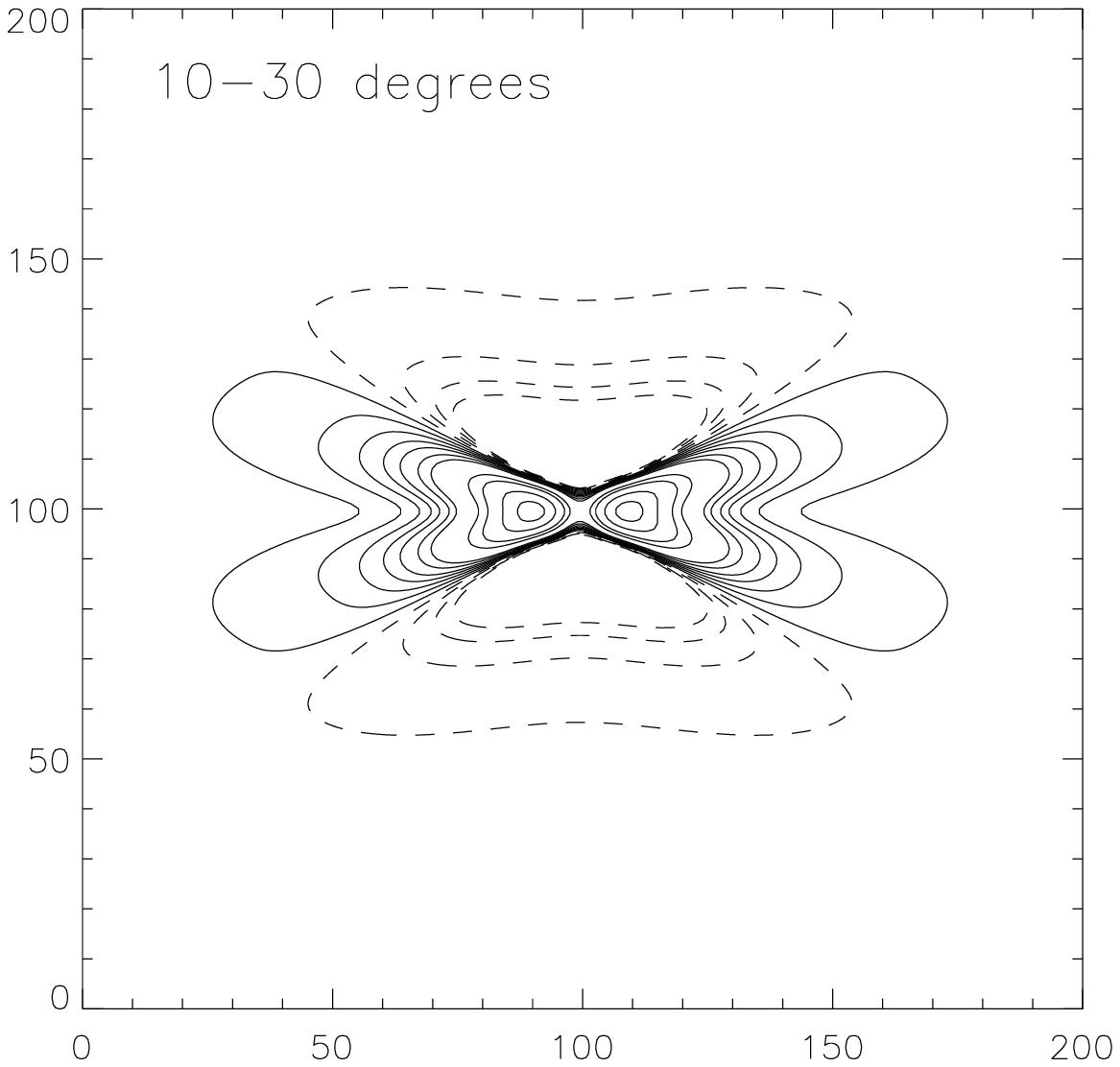}

\plottwo{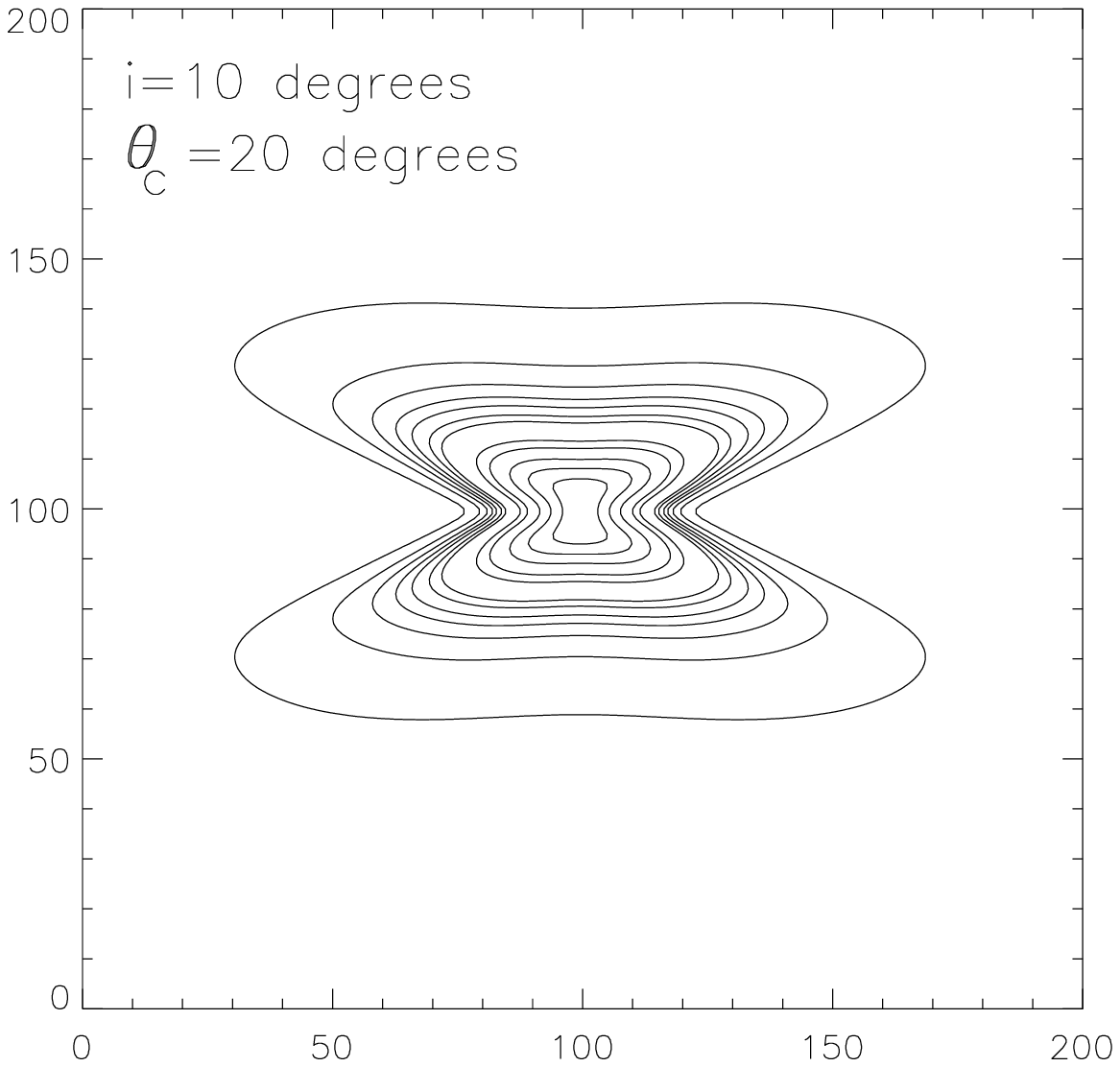}{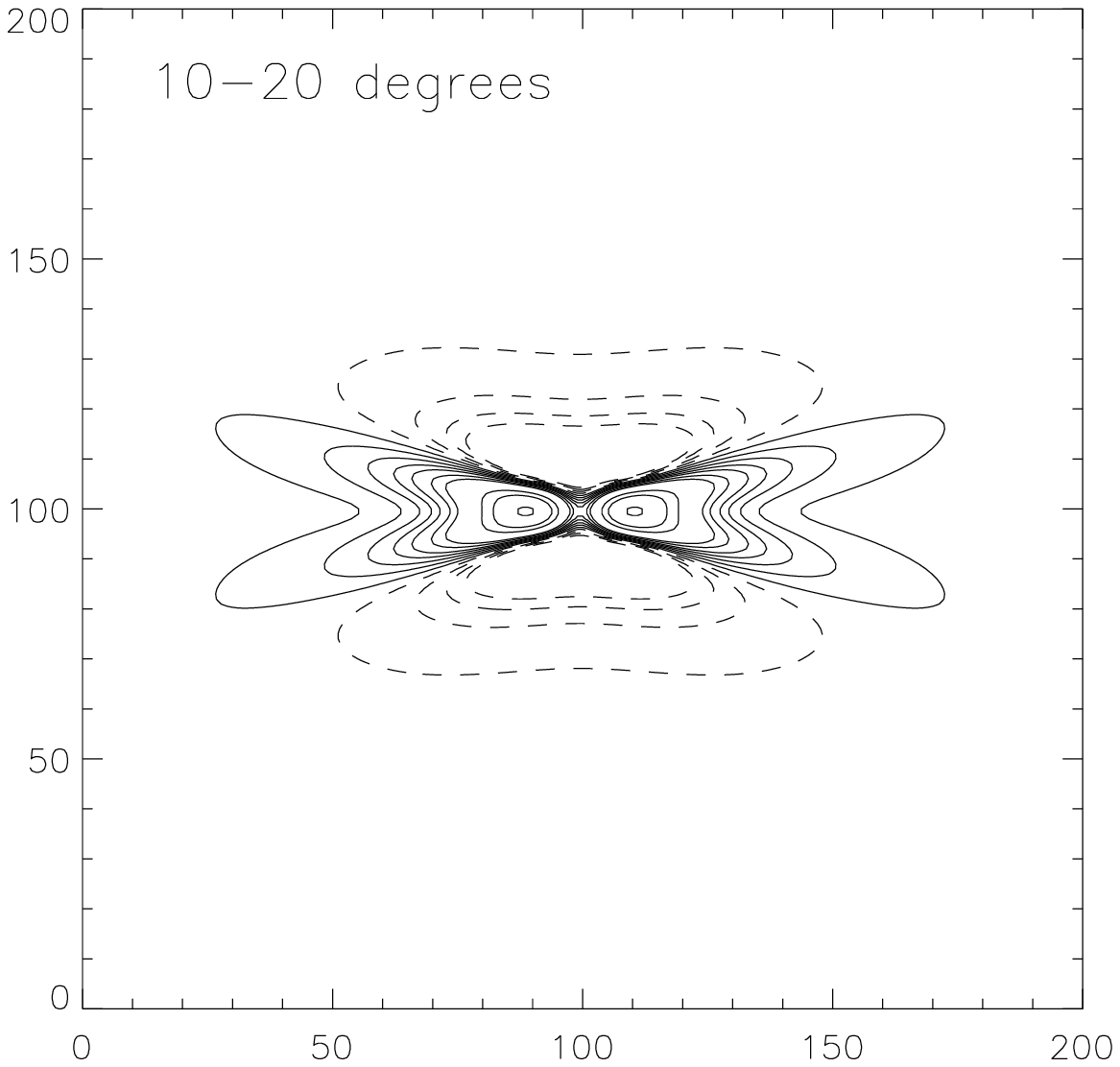}

\plottwo{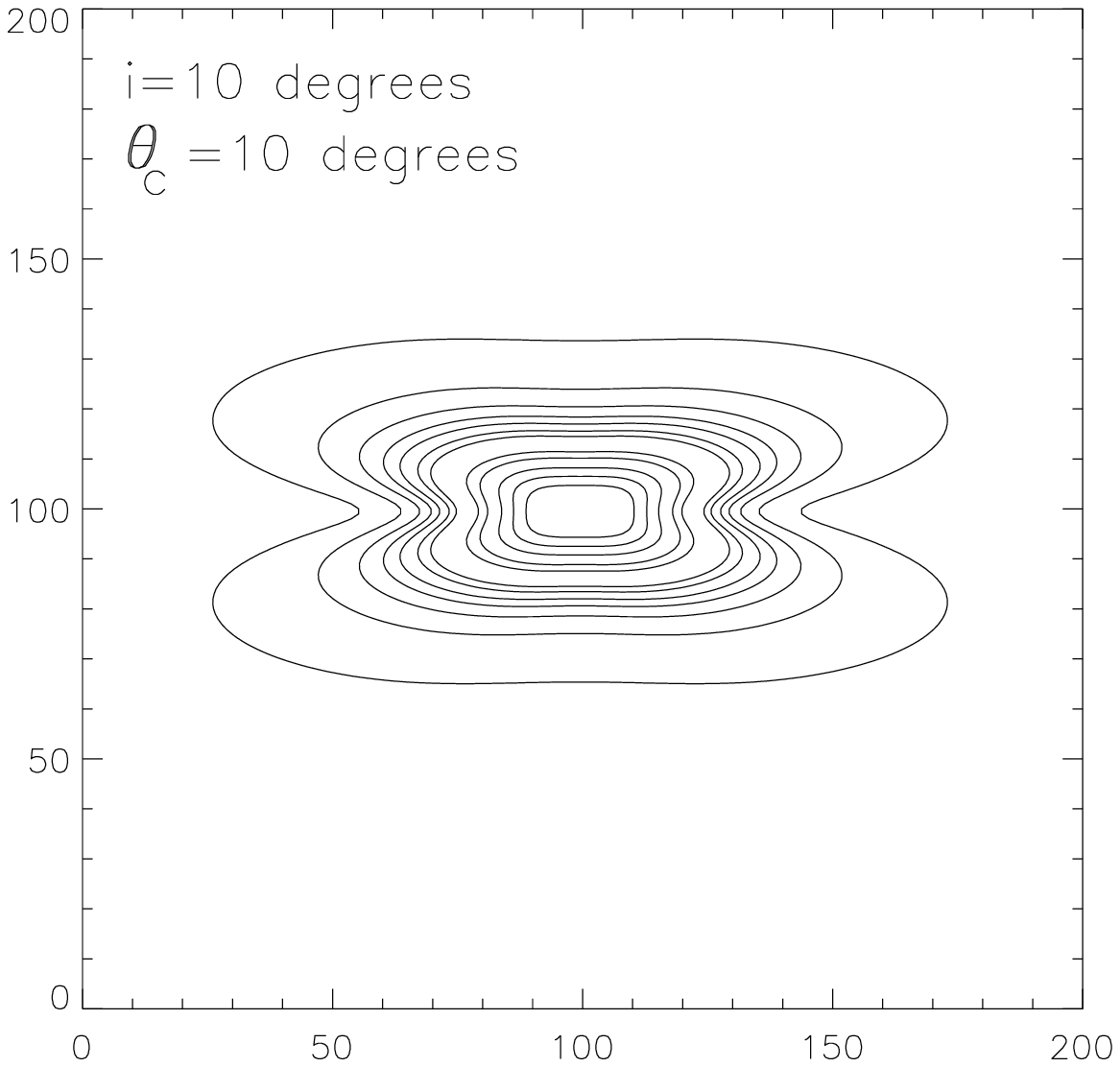}{f6f.eps}

\end{center}

\caption{(Left) Contour images of the model with variable
critical angle $\theta_c$ described in the text.
The inclination angle of the plane of the disk of the model with respect to
the line of sight is always $10^\circ$ and the critical angles are
indicated in each figure.
(Right) Contour plots of the model difference images.
The differences refer to the critical angle $\theta_c$.
Contours are
-100,-80, -60, -40, -30, -15, -12, -9, -7, -5, -3, -1,
1, 3, 5, 7, 9, 12, 15, 30, 40, 60, 80, and  100
percent the peak value of the model with
critical angle of $30^\circ$.
The $x$ and $y$ axes of the models are in units of $0\rlap.{''}01$ and the images are smoothed to an
angular resolution of $0\rlap.{''}1$.
\label{fig6}}
\end{figure}

\clearpage

\begin{deluxetable}{lcc}
\tablewidth{12.0cm}
\tablecaption{Configuration A Observations at 2 cm Analyzed}
\tablehead{
\colhead{}  & \colhead{Phase} 
& \colhead{Synthesized Beam} \\  
\colhead{Epoch}  & \colhead{Calibrator} 
& \colhead{($\theta_M \times \theta_m; PA$)\tablenotemark{a}} \\
}
\startdata
1983 Oct 30 (1983.83) & 2007+404 & $0\rlap.{''}12 \times 0\rlap.{''}10;~+5^\circ$ \\
1995 Aug 24 (1995.65) & 2007+404 & $0\rlap.{''}13 \times 0\rlap.{''}11;~+30^\circ$ \\ 
1996 Dec 05 (1996.93) & 2007+404 & $0\rlap.{''}12 \times 0\rlap.{''}11;~-32^\circ$ \\ 
2004 Nov 15 (2004.87) & 2007+404 & $0\rlap.{''}12 \times 0\rlap.{''}11;~+89^\circ$ \\
\enddata
\tablenotetext{a}{Major axis $\times$ minor axis in arcsec; PA in degrees for synthesized
beam made with ROBUST = 0.}

\end{deluxetable}

\clearpage

\begin{deluxetable}{lccc}
\tablewidth{15.0cm}
\tablecaption{Flux Densities from Configuration D Observations at 2 cm}
\tablehead{
\colhead{}  & \colhead{Phase}
& \colhead{Synthesized Beam} & \colhead{Flux Density} \\
\colhead{Epoch}  & \colhead{Calibrator}
& \colhead{($\theta_M \times \theta_m; PA$)\tablenotemark{a}} & \colhead{of MWC~349A (mJy)} \\
}
\startdata
1984 Aug 19 (1984.63) & 2007+404 & $4\rlap.{''}3 \times 3\rlap.{''}9;~-89^\circ$ & 390$\pm$30 \\
1996 Jul 30 (1996.58) & 2007+404 & $4\rlap.{''}4 \times 3\rlap.{''}8;~+11^\circ$ & 383$\pm$20 \\
2001 Nov 09 (2001.86) & --------\tablenotemark{b} & $5\rlap.{''}3 \times 4\rlap.{''}3;~-90^\circ$ & 380$\pm$6 \\
2006 Jan 20 (2006.05) & --------\tablenotemark{b} & $9\rlap.{''}4 \times 5\rlap.{''}4;~-55^\circ$ & 378$\pm$15 \\ 

\enddata
\tablenotetext{a}{Major axis $\times$ minor axis in arcsec; PA in degrees for synthesized
beam made with ROBUST = 0.}
\tablenotetext{b}{These observations were made without a phase calibrator and
the flux calibration was transferred directly from 1331+305, the amplitude
calibrator, to MWC~349A.}

\end{deluxetable}

\clearpage

\begin{deluxetable}{lcc}
\tablewidth{12.0cm}
\tablecaption{Configuration A Observations at 1.3 cm Analyzed}
\tablehead{
\colhead{}  & \colhead{Phase} 
& \colhead{Synthesized Beam} \\  
\colhead{Epoch}  & \colhead{Calibrator} 
& \colhead{($\theta_M \times \theta_m; PA$)\tablenotemark{a}} \\
}
\startdata
1985 Jan 20 (1985.05) & 2007+404 & $0\rlap.{''}08 \times 0\rlap.{''}07;~-22^\circ$ \\
1990 Mar 29 (1990.24) & 2007+404 & $0\rlap.{''}09 \times 0\rlap.{''}08;~-68^\circ$ \\ 
2004 Nov 15 (2004.87) & 2007+404 & $0\rlap.{''}09 \times 0\rlap.{''}07;~-85^\circ$ \\
\enddata
\tablenotetext{a}{Major axis $\times$ minor axis in arcsec; PA in degrees for synthesized
beam made with ROBUST = 0.}

\end{deluxetable}

\clearpage

\begin{deluxetable}{lcc}
\tablewidth{12.0cm}
\tablecaption{Configuration A Observations at 6 cm Analyzed}
\tablehead{
\colhead{}  & \colhead{Phase}
& \colhead{Synthesized Beam} \\
\colhead{Epoch}  & \colhead{Calibrator}
& \colhead{($\theta_M \times \theta_m; PA$)\tablenotemark{a}} \\
}
\startdata
1982 Jun 04 (1982.42) & 2007+404 & $0\rlap.{''}37 \times 0\rlap.{''}34;~+2^\circ$ \\
1996 Dec 05 (1996.93) & 2007+404 & $0\rlap.{''}38 \times 0\rlap.{''}34;~-9^\circ$ \\
2004 Nov 15 (2004.87) & 2007+404 & $0\rlap.{''}38 \times 0\rlap.{''}34;~-80^\circ$ \\
\enddata
\tablenotetext{a}{Major axis $\times$ minor axis in arcsec; PA in degrees for synthesized
beam made with ROBUST = 0.}

\end{deluxetable}

\end{document}